# A Meta-Fusion Architecture for Few-Shot Classification of Spike Waveforms in High-Bandwidth Brain-Machine Interfacing

Tao Fang, *Student Member, IEEE*, Majid Zamani, *Member, IEEE*

***Abstract*: There is a need for fast adaptation in spike sorting algorithms to implement brain-machine interface (BMIs) in different applications. Learning and adapting the functionality of the sorting process in real-time can significantly improve the performance. However, deep neural networks (DNNs) depend on large amounts of data for training models and their performance sustainability decreases when data is limited. Inspired by meta-learning, this paper proposes a few-shot spike sorting (FS-SS) framework with variable network model size that requires minimal learning training and supervision. The framework is not only compatible with few-shot adaptations, but also it uses attention mechanism and dilated convolutional neural networks. This allows scaling the network parameters to learn the important features of spike signals and to quickly generalize the learning ability to new spike waveforms in recording channels after few observations. The FS-SS was evaluated by using freely accessible datasets, also compared with the other state-of-the-art algorithms. The average classification accuracy of the proposed method is 99.28%, which shows extreme robustness to background noise and similarity of the spike waveforms. When the number of training samples is reduced by 90%, the parameter scale is reduced by 68.2%, while the accuracy only decreased by 0.55%. The paper also visualizes the model's attention distribution under spike sorting tasks of different difficulty levels. The attention distribution results show that the proposed model has clear interpretability and high robustness.**

## I. INTRIDUCTION

Extracellular recordings have been widely used to monitor neuronal activity by implanting multi-electrodes in the cortex and capturing multidimensional neural data. A processing step, known as spike sorting shown in Fig.1, is necessary to separate the multi-unit neural activities and assign the captured spikes to their originating neurons [1-7]. Spike sorting is an invaluable research tool applied in implantable brain-machine interface (iBMI) research for studying and decoding neural signals from different brain regions and understanding the mechanisms of the brain. It is extremely beneficial in design and development of various applications such as identifying the optimal patterns and parameters to condition diseases by artificially modulating irregular or faulty electrical impulses [8], realizing a communication bridge for control of assistive devices for patients with damaged sensory/motor functions such as hand prosthesis [9], and stimulating a particular pathway for biological functionality regularization [10].

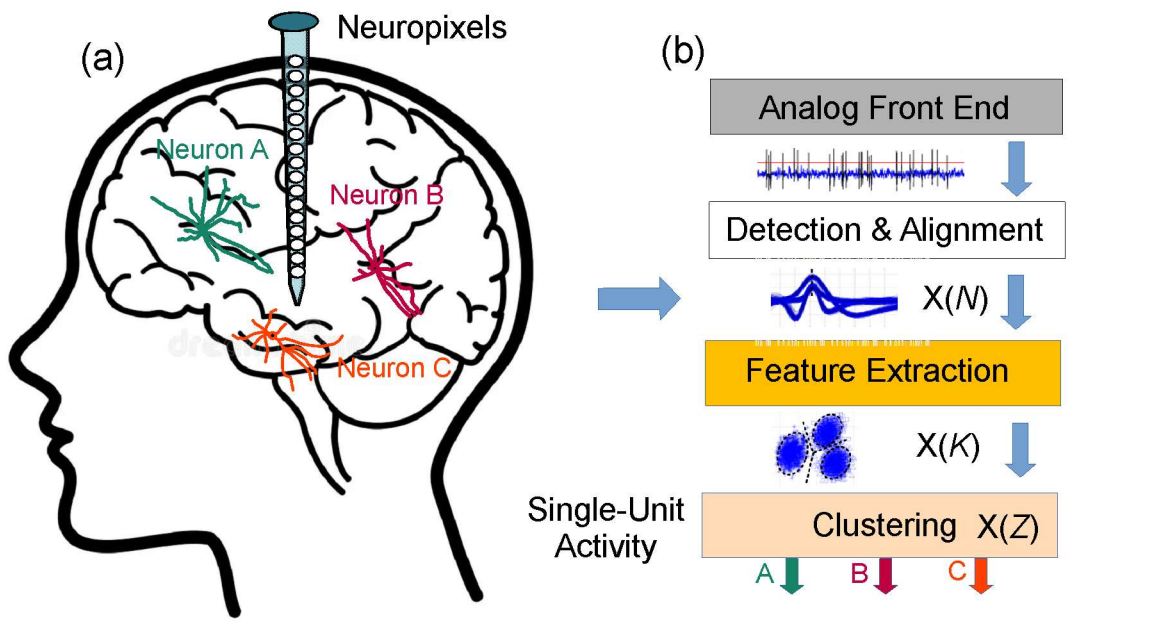

Fig. 1. (a) Recording activities from neurons A, B and C using Neuropixels electrode arrays. (b) Spike sorting process for determining single unit activity. Data is progressively reduced over processing blocks (Z <K< N).

The recent trend in brain sensing is about the utilization of high-channel count neural interfaces that include tens of thousands of sensing probes [11]. There are two important aspects associated with the large-scale data streaming:

(i) *Recording channel variability:* three factors are involved: a) neurons can show different dynamics based on electrode location. (b) neurons have various structures depending on the cortex regions and layers [12] which results in generation of unique actions potentials. And (c) action potentials are also function of time [13]. Considering such variations, there is high probability of spike shape changes over time. Therefore, a solid model is required to adapt itself to the recording channel variations.

(ii) *Training time:* The second challenge is training time which means adaptation to the characteristics of the recording channel. This certainly requires huge amount of training data as the deep learning algorithms are data hungry [14].

There is a need for a spike classification algorithm that adapts (or learns), and embeds intermittently the information about

T. F. Author was with Engineering and Applied Technology Research Institute, Fudan University, 220 Handan Road, Shanghai, 200433, China (e-mail: 19210860039@fudan.edu.cn).

M. Z. Author is with the School of Electronics and Computer Science, University of Southampton, Southampton, SO17 1BJ UK, (e-mail: m.zamani@soton.ac.uk).

new channel with only a few observations through a fast learning approach.

In recent years, different types of DNNs have been developed and utilized in spike sorting [15]. However, few-shot learning is not widely adopted in spike sorting. For example, Wu et al. [16] in 2019 developed a semi-supervised spike classification method called few-shot spike sorting using general adversarial network (GAN). The authors in [16] successfully categorized spikes within the utilized synthetic data: using 50 labelled spikes from a recording with a noise standard deviation of 5%, an accuracy of 97.7% was achieved by increasing the number of labelled spikes to 200, the accuracy improved to 98.2%. In the presence of noise with a standard deviation of 20%, they reached an accuracy level of 66.4% (50 labelled spikes) and 85.5% (200 labelled spikes) [16]. Also tested on experimental data, it was demonstrated that training the GAN on particular waveforms with only a "few shots" was sufficient to recognize similar spikes in inference mode.

Few-shot learning is also used in other biomedical signal processing applications. In [17], authors proposed a novel framework referred to as the few-shot hand gesture recognition (FS-HGR). The FS-HGR quickly adapts by using a small number of observations to a new gesture (or user) through learning on its prior experience. Tested on second Ninapro databases (also referred to as the DB2) [18], which consists of 50 gestures (rest included) from 40 healthy subjects, FS-HGR led to 85.94% classification accuracy. FS-HGR led to 83.99% classification accuracy on new repetitions with few-shot observations (5-way 5-shot). Also, authors in [17] achieved 81.29% accuracy on new subjects with few-shot observation (5-way 5-shot). In [19], authors introduced a few-shot learning (FSL) applicability for electrocardiogram (ECG) signal proximity-based classification. They embedded the FLS to a deep convolutional neural network to recognize 2, 5 and 20 different heart disease classes. The extracted QRS complexes from PTB-XL dataset [20] containing the labelled 10-second raw signal ECG was used to train the model in [19]. The FSL-based classification provides continuous classification accuracy augmentation without network adjustments, and it achieves classification of healthy and sick patients ranging from 93.2% to 89.2%.

Additionally, spiking neural networks (SNNs) inspired by biological brains have emerged as a promising alternative for spike sorting due to their event-driven nature and inherent energy efficiency [21]. For example, the work in [22] presents an FPGA-implemented of an unsupervised SNN for real-time, low-latency, and power-efficient multi-channel spike sorting. The [23] proposes a low-power, variation-tolerant and adaptive, spike sorter for neural implant devices, leveraging a two-step supervised on-chip training module and a resistive crossbar SNN classifier for real-time spike classification on FPGAs. However, despite these advantages, training SNNs to achieve high accuracy and robustness comparable to conventional deep neural networks (DNNs) on complex, noisy spike data remains a significant challenge, often requiring sophisticated learning rules or larger datasets for generalization [24]. Furthermore, while SNNs naturally align with the temporal sparsity of neural signals, their application to few-shot learning scenarios in spike sorting is still an evolving area, demanding innovative approaches to truly leverage their strengths in data-limited settings.

In essence, a solid and structured study on few-shot learning for spike sorting application is required, as having access to an implantable processor for changing its parameters is an extremely challenging and tedious task. This paper introduces a few-shot learning framework for improved spike classification accuracy. Compared with traditional methods, the introduced framework only requires a small number of training data to achieve high-performance deep learning spike sorting. The main contributions of this paper are summarized as follows:

- The FS-SS framework alleviates the problem of few observations learning in the spike sorting where the number of training samples are extremely limited. This makes BMI systems quicker to set up and more adaptable in real-time.
- With a variable network model size, FS-SS dynamically adjusts its complexity, leading to substantial reductions in parameters and computational overhead (e.g., 68.2% reduction when training data is reduced by 90%). This is appealing for resource-limited BMI hardware, enabling faster processing and lower energy use.
- Unique building blocks are designed and introduced in FS-SS. Embedding module to integrate few-shot adaptations capability, residual attention (RA) module to enhance classification robustness in noisy conditions and residual dilated convolution (RDC) to capture neuronal activity patterns at different scales and improve the model's ability to process multi-scale features while maintaining suitable resolution and reducing computational complexity.
- FS-SS employs meta-learning, allowing the model to "learn how to learn." This enables rapid generalization to new spike waveforms and unseen recording channels with minimal samples, drastically reducing the need for extensive retraining in large-scale BMI deployments.

To the best of our knowledge, this is the first time that a paper outlines in detail a few-shot adaptation algorithm in spike sorting. The rest of the paper is structured as follows: Section II describes the proposed architecture, including details of the individual constituent modules and how these modules are organized in the FS-SS framework. Section III describes the used datasets for performance analysis and the evaluation methods. Section IV scrutinizes the performance of the FS-SS framework by comparing it with other sorting methods, testing adaptation with different dataset proportions and visualization of sorting and attention results, and ablation experiments were performed to explore the impact of a single module on performance. Finally, Section V makes some concluding remarks.

## II. Proposed Spike Sorting Architecture

This section first briefly explains the meta-learning method for the few-shot problem, then describes the basic building blocks of the proposed FS-SS network, and finally illustrates the overall FS-SS architecture and how to align the meta-learning method with the spike sorting process.

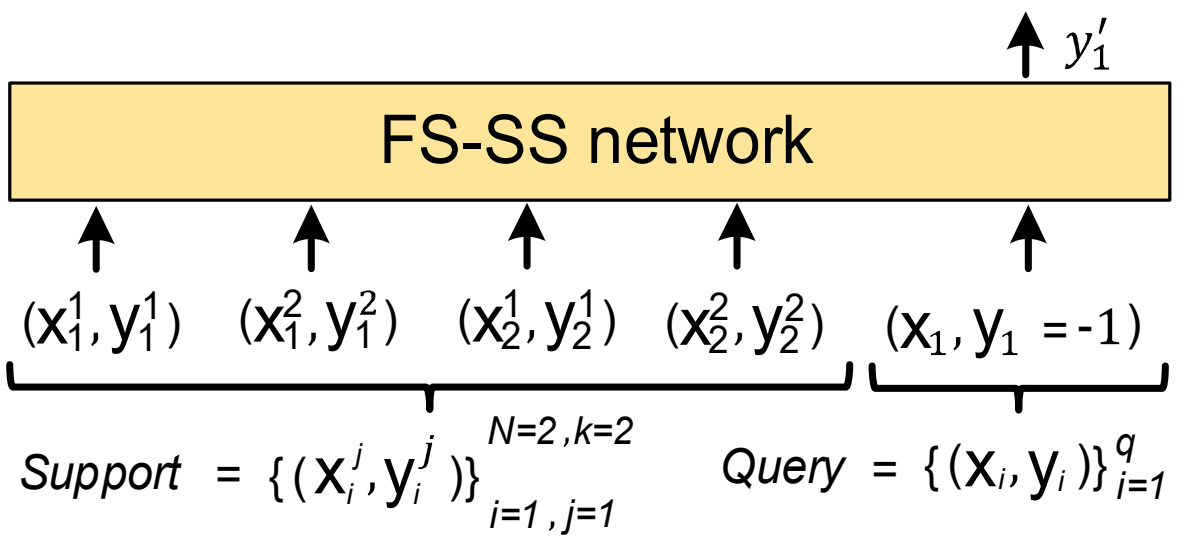

Fig.2. Meta-learning-based FS-SS network training process. The input includes support spikes and query spikes. In this paper, the support set contains two classes ($N$=2), each class includes two spikes ($k$=2), and the query set contains only one spike ($q$=1). The FS-SS network is used to predict the label corresponding to the query set spike.

## A. Meta-learning Problem

In recent years, meta-learning [25] has been offering an extremely effective approach in solving problems that require minimal supervision [26-28]. Meta-learning is also called learning to learn aiming to improve the adaptability of the model to new tasks by learning the distribution of multiple tasks, rather than relying solely on the data of a single task [29]. This learning approach enables the deep learning algorithms to quickly adapt to new tasks by training on multiple different but related tasks. It is extremely suitable for classification problems with high complexity, high dimensions, and few samples such as spike sorting.

Unlike traditional supervised learning methods, the minimum data for updating neural network parameters based on meta-learning is a task rather than a single spike sample [30]. A task contains multiple spikes, including support ($S$) spikes and query ($Q$) spikes, so a task consists of $S + Q$ spikes. The support set and query set are defined as in Eq. (1) and (2), where $x_i^j$ represents the j[th] sample of the i[th] class and $y_i^j$ is its corresponding label. The query set usually has only one sample per class, so the superscript j is omitted.

$$S = \{(x_i^j, y_i^j)\}_{i=1,j=1}^{N,k} \quad (1)$$

$$Q = \{(x_i, y_i)\}_{i=1}^{q} \quad (2)$$

The meta-learning problem with $N$ classes and $k$ samples in each category is called $N$-way $k$-shot problem [30] as shown in Fig. 2. Given the datasets contains limited categories (3-class and 6-class scenarios). Choosing 2-way 2-shot configuration, avoiding potential class imbalance during sampling. Which means that the support set has 2 classes ($N$=2) each class has 2 samples ($k$=2) with one sample in the query set ($q$=1). Therefore, a total number of 5 spike waveforms, which can be expressed as follows:

$$S = \{(x_1^1, y_1^1), (x_1^2, y_1^2), (x_2^1, y_2^1), (x_2^2, y_2^2)\} \quad (3)$$

$$Q = \{(x_1, y_1)\} \quad (4)$$

Inspired by [31], the true labels of the support set are fed into the network while the labels of the query set are treated as non-existent labels represented by $-1$ in the training phase. Then these $N \times k + 1$ labels and their corresponding sequences X are fed into the network. In the parameter update phase, the loss value $\mathcal{L}_Q(f_\theta)$ is calculated only for the output label of the query set $f_\theta(x_q)$ and the true label $y_q$ as indicated in Eq. (5). This loss is used to update the learnable parameters $\theta$ of the entire network and obtain the new parameters $\theta'$. Where $\alpha$ represents the learning rate, and $\nabla_\theta$ represents the result of the gradient of $\theta$ as indicated in Eq. (6). After multiple iterations of the above process, the model will learn how to encode the support set and predict the true label of the query set [28].

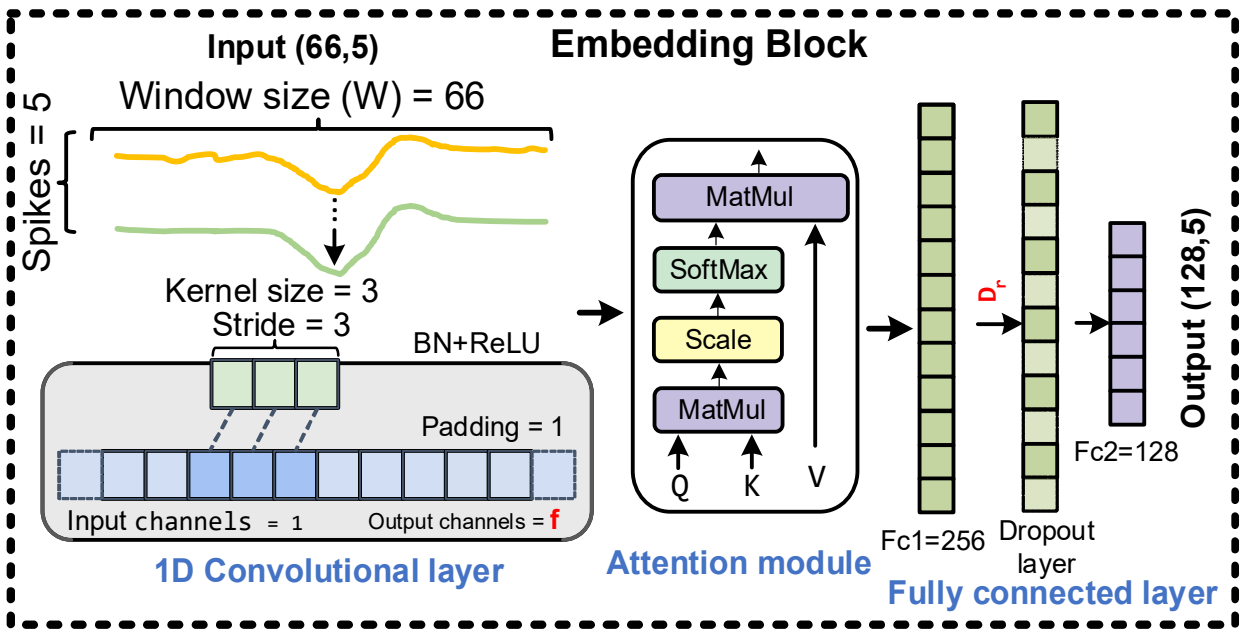

Fig.3. Details of embedding module of the FS-SS network. This module is used to standardize the input sequence, and also facilitates the use of few-shot capability in the deep learning processing pipeline. It includes a 1D convolution layer with a variable number of output channels $f$ depending on the size of the dataset, a batch normalization layer (BN), a self-attention mechanism unit and two fully connected layers (Fc1 and Fc2) outputs a 128-dimensional feature vector.

$$\mathcal{L}_Q(f_\theta) = loss(f_\theta(x_q), y_q) \quad (5)$$

$$\theta' = \theta - \alpha \nabla_\theta \mathcal{L}_Q(f_\theta) \quad (6)$$

## B. The Building Modules of the FS-SS Framework

This section explains the design ideas and implementation details of the constituent modules in the FS-SS framework. It finally outlines the overall architecture of the FS-SS.

### 1) The Embedding Module

The embedding module integrates the few-shot observations capability onto the FS-SS network. It performs preliminary feature extraction and to map input spike waveforms (5 spikes, 66 samples per spike) onto a standardized feature vector [32]. As shown in Fig.3, the embedding module first extracts the primary time domain features of the spike signal through a 1D convolution layer. The kernel size in the 1D convolution layer is set to 3, it has stride of 1 and also padding size is set to 1. The number of output channels shown by $f$ in the 1D convolution layer (i.e., the number of convolution kernels) is not fixed and it depends on the number of the spikes used in the training dataset. When the training dataset is large, the number of convolution kernels can be dynamically increased to capture wider features. On the other hand, when the training dataset is small, the number of convolution kernels can be appropriately reduced to avoid overfitting. This design ensures that the model can dynamically adjust its complexity according to the size of the dataset, thereby achieving a balance between extracting rich features and preventing overfitting. In addition, smaller number of parameters means less computation and lower computation cost which is more suited more to implantable brain processing. A batch normalization (BN) layer is connected after the convolution layer to speed up the training process [33].

The number of output channels $f$ of the 1D convolutional layer shown in the embedding module in Fig.3 is calculated by linear interpolation according to the size of the $N_d$, where $N_d$ represents the number of spikes used to train the model. Assuming that the maximum number of convolution kernels is $f_{max}$, the minimum number of convolution kernels is $f_{min}$, the maximum dataset size is $N_{max}$, and the minimum dataset size is $N_{min}$, the basic steps of the process are as follows:
a) Determine the normalization ratio $P_{std}$ used for the dataset size:

$$P_{std} = \frac{N_d - N_{min}}{N_{max} - N_{min}} \quad (7)$$

b) Then, the initial value of the number of convolution kernels $f_{linear}$ is calculated by linear interpolation:

$$f_{linear} = f_{min} + \left(P_{std} \times (f_{max} - f_{min})\right) \quad (8)$$

c) Finally, $f_{linear}$ is rounded to the nearest power of 2 to get $f$, where $\lfloor\cdot\rfloor$ represents rounding in Eq. (9). Adjusting the number of convolution kernels to a power of 2 can better utilize the memory alignment and parallel computing capabilities of the GPU:

$$f = 2^{\lfloor \log_2(f_{linear}) + 0.5 \rfloor} \quad (9)$$

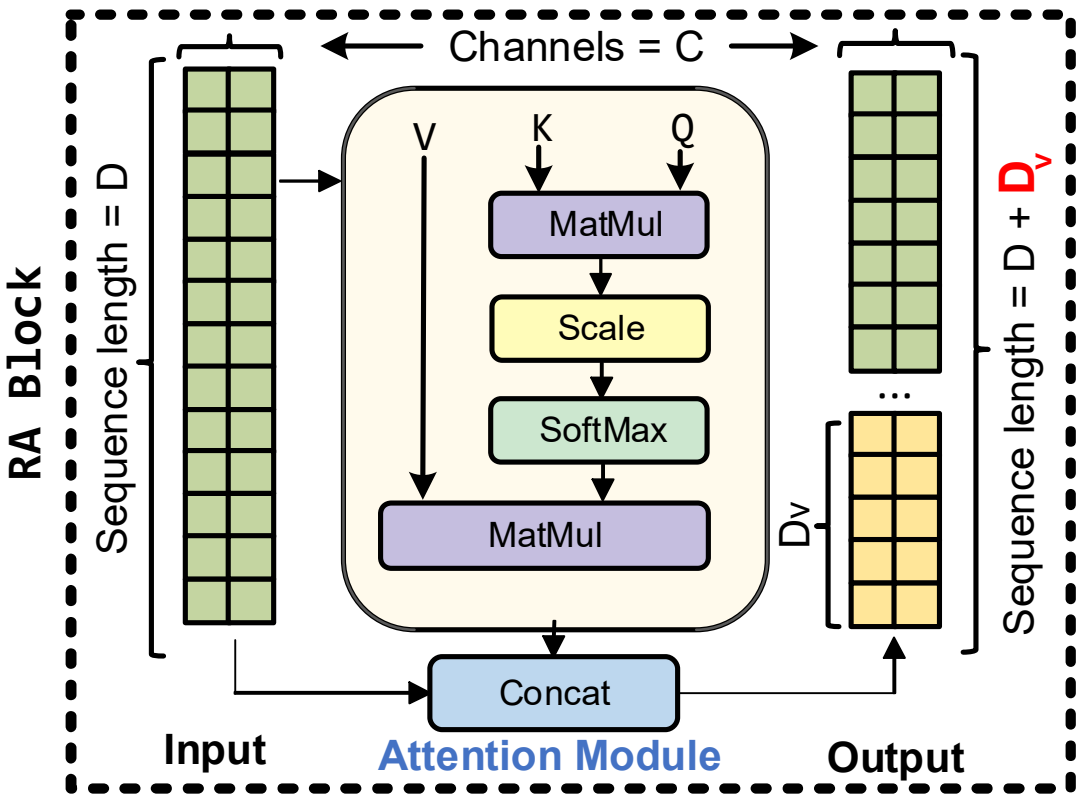

Fig.4. The RA block of the FS-SS network. The feature matrix of C×D is linearly mapped using three fully connected layers to key (K), query (Q) and value (V). The length of processed input data D is mapped to D + Dv where Dv corresponds to the additional length introduced by the attention mechanism.

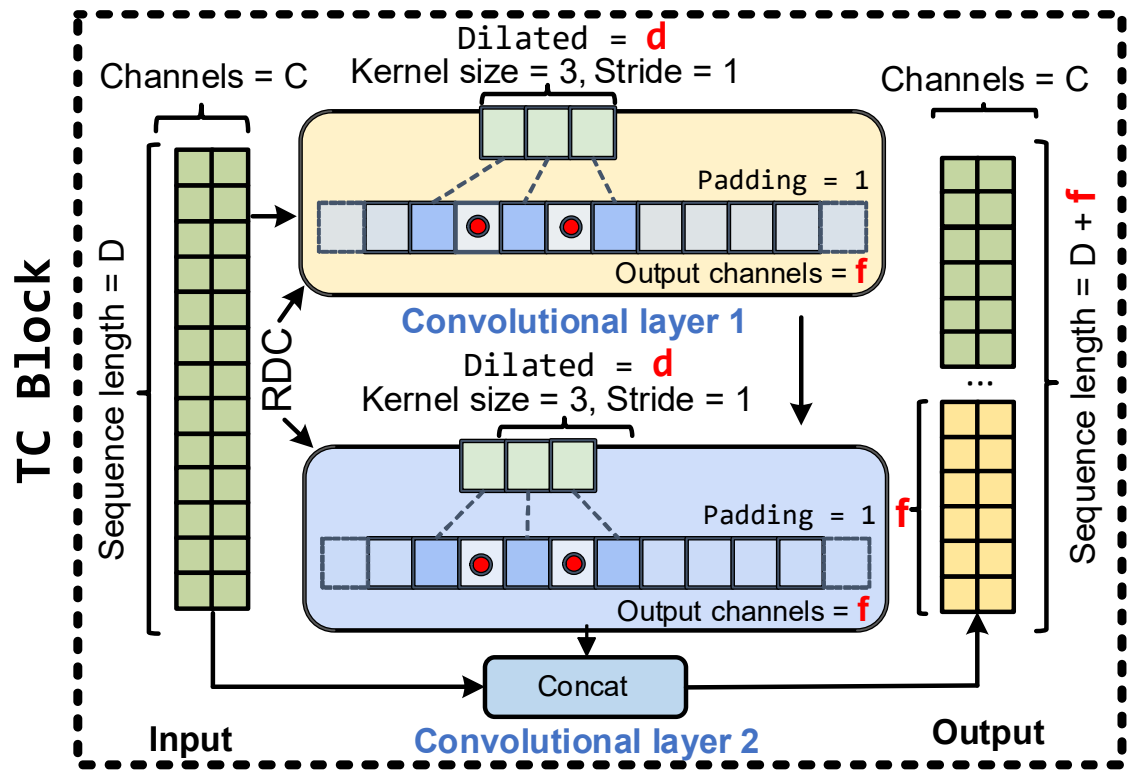

Fig.5. The RDC block used in the FS-SS network consists of two dilated convolutional layers (Layer 1 and Layer 2) and a residual operation, where the dilation rate (d) and the number of convolutional layer output channels are variable. Details of RDC are annotated on figure. TDC maps C×D to an output feature vector of C×(D+f).

The determination of $f$ is performed in model initialization phase and is utilized in the embedding module during the training phase. A basic self-attention unit follows the convolutional layer, taking the output of the previous convolution layer as input and further processing these features. Specifically, by adding the self-attention mechanism unit, the embedding module can generate the corresponding attention weight matrix by learning the primary time domain features obtained by convolution, so that the FS-SS model can focus more on the important time domain features of the spike signal. Finally, the output of the attention unit passes through two fully connected layers shown by Fc1 and Fc2 in Fig.3. The number of neurons in the first fully connected layer depends on the output feature dimensions of the attention unit after flattening, which is also closely related to the number of the kernels in the convolutional layer (because the input and output dimensions of the attention module are unchanged), which means that the number of neurons in this layer is also adaptive. After the first fully connected layer (Fc1), a dropout layer is connected to prevent the "co-adaptation" problem [34]. The dropout rate is also determined by the size of the $N_d$, and linear interpolation is also used. In this work, the maximum dropout rate is 0.5 and the minimum dropout rate is 0.1. The dropout rate is inversely proportional to the size of the dataset, because small-scale datasets are prone to overfitting, so a larger dropout rate is required [34]. The output of the second fully connected layer (Fc2) is fixed to 128 to standardize the embedding results. This means that no matter what the scale of the input data is, the final embedding result length is always 128, thus ensuring the consistency of subsequent processing.

*2) Residual Attention (RA) Block*

Residual attention (RA) block based on the self-attention mechanism is considered in the network to capture the potential relationships of the features from the embedding module, and to provide more abstract sequence of features. The attention mechanism has been proven to be very effective in processing features with long-distance dependencies [32] so it is natural to apply the attention mechanism for tasks with multi-scale temporal features such as spike sorting shown in Fig.1.

The input of the RA accepts $C$ channels with the input length of $D$, which corresponds to the number of extracted feature vectors from the previous stage. For example, the input of the first RA ($C$=128, $D$=5) comes from the embedding module consisting of extracted features from 4 support spike waveforms and 1 query spike.

The number of channels processed by RA is $C$, so $C$ will not change before and after passing through this module. The feature matrix of $D \times C$ is linearly mapped using three fully connected layers to key ($K$), query ($Q$) and value ($V$). The corresponding fully connected layer sizes are recorded as $D_Q$, $D_K$ and $D_V$, and the corresponding weight matrices are called $W_Q$, $W_K$ and $W_V$ respectively. The attention score is then calculated by the dot product between $K$ and $Q$ which reflects the dependencies between the features. Subsequently, in order to prevent the gradient saturation caused by excessive calculation results, the dot product result of $K$ and $Q$ is divided by the scaling factor $\sqrt{D_K}$. The Softmax function is then applied to convert the calculation result into a weight, and finally the weight is applied to the value vector $V$ to obtain the corresponding attention output. The whole process is expressed as:

$$Attention(Q, K, V) = softmax\left(\frac{QK^T}{\sqrt{D_K}}\right)V \quad (10)$$

Finally, in order to prevent degrading the performance of the model from due to the deepening of the network layers, residual connection is used in the RA block. The feature with a length of $D$ becomes $D_V$ after being processed by the attention mechanism (Eq. 10), and the original features and processed features are concatenated, so the final output feature length of the RA block is $D + D_V$.

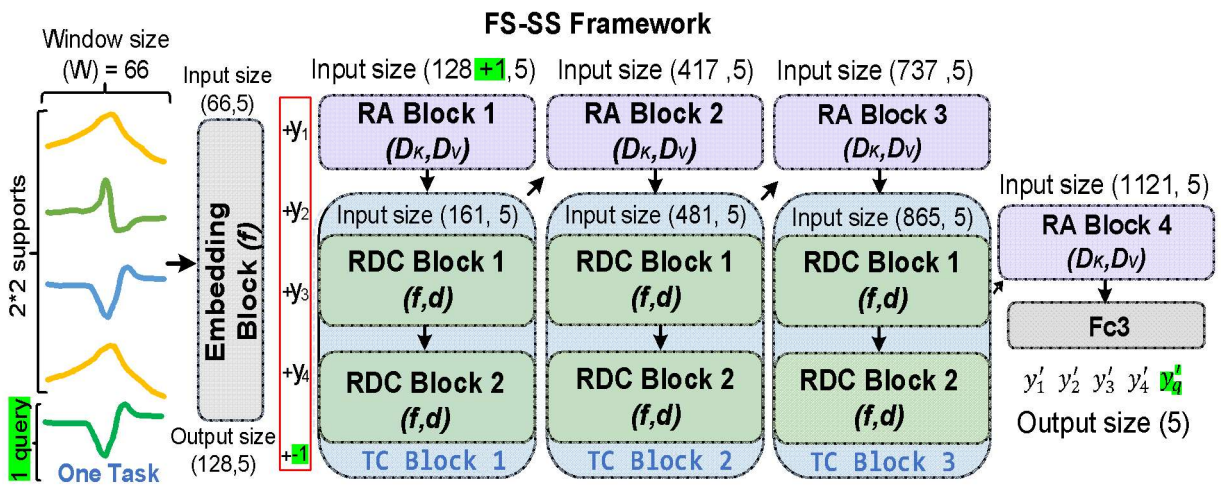

Fig.6. Overall architecture of the proposed FS-SS network. The network consists of an embedding module, RA blocks and three TC modules. Each TC module stacks two RDC blocks. The figure shows the input and output dimensions of each module when the input spike data (66,5). 66 is the length of the extracted spike waveform, and 5 represents 4 support set spikes and one query set spike. The dilation factors in the TDC units increase exponentially for effective expansion of the receptive field without introducing significant computation cost.

*3) Residual Dilated Convolutions (RDC) Block*

Inspired by [31] and [35], a RDC block is designed that consists of two dilated convolutional layers and a residual operation as shown in Fig.5. The RDC block extracts more abstract features of the spike waveforms and it achieves a wider receptive field by sequentially stacking more RDC blocks.

Traditional convolution can be used to achieve a wider receptive field. However this needs to expand the size of the convolution kernels or stack more convolution layers [36]. Both methods inevitably introduce a sharp increase in computational complexity, limiting the practical use of deep neural networks in iBMIs. Therefore, dilation convolution is utilized in FS-SS to increase the receptive field of the convolution kernel without increasing the number of parameters (i.e. the implementation cost). This convolution method has significant advantages in processing multi-scale features and maintaining spatial resolution [37].

Dilated convolution significantly increases the receptive field without increasing the number of parameters [37] by scaling the convolution kernels as shown by red circles in Fig.5. Utilizing scaling factors, a spike waveform is decomposed to different time features. A temporal convolution (TC) module consists of two RDCs as shown in Fig.5. The TC modules are then stacked multiple times to achieve gradual increase of receptive field. Therefore, a smaller dilation rate is used to capture short-term peak features in the first RDC module. For instance, the first RDC has the dilation rate of 1, its corresponding convolution kernel size is set to 3 with a stride set to 1. Also, the padding in the first RDC is set to 1 to keep the input and output dimensions equal. For the subsequent RDC modules, larger dilation rates need to be used to obtain deeper receptive filed. Following [37] and [38], the dilation rate of the RDC modules is set to twice that of the previous RDC module to capture the longer features of the spike signal. Therefore, the dilation rate corresponding to two sequentially connected RDC modules increases exponentially which can effectively expand the receptive field. The number of RDC modules ($Z$) in a TC depends on the complexity of the meta-learning task, which is reflected here as the sum of the number of spikes in the support set and the query set $S + Q$ determined by the following equation:

$$Z = [log_2(S + Q)] \qquad (11)$$

where $[\cdot]$ represents rounding operation. As the RDC modules are stacked, the neural network will learn more and more abstract features. However, this will result in a serious gradient vanishing problem. As expressed in Eq. (12), the residual mechanism ensures that the gradient does not disappear during the backpropagation process by directly passing information to the subsequent layers [39]. For a input feature vector $(x)$ with a length of $D$, the dimension is changed to $D + f$ after being processed by TC due to the residual operation, where $f$ is determined by the number of convolution kernels. This allows to build and stack more RDC modules for implementing a deeper network.

$$RA\ output = x + f(x, \theta) \qquad (12)$$

## C. Architecture of FS-SS

As shown in Fig.6, the entire FS-SS network consists of 1 embedding module, 4 RA blocks and 3 TC modules. Each TC module consists of 2 RDC blocks. Finally, a basic fully connected layer (Fc3) is added to predict the spike classes. Fig.6 also shows the dimensional changes of the input samples after passing through each module under a task. In the proposed FS-SS network, 4 support spikes are included in the support set (i.e. 2-way 2-shot), and the query set consists of 1 spike highlighted in green. All 5 spikes are sent to the embedding module, and the final output feature vector dimension is standardized to 128 (128, 5). Subsequently, the true label of the support set is added to the 128-dimensional feature as an additional feature, and the label of the query set is blurred to a non-existent label -1 [31]. The concatenated 129-dimensional features are sent to the first RA block to learn the attention relationships between features. The attention results are concatenated with the residual path, and the final output dimension is 129+32 features where 32 represents the size of the $D_V$ of the RA1 block. In the next three RA blocks (RA2, RA3 and RA4), due to the continuous increase in feature dimensions, the corresponding attention range also needs to be expanded accordingly. So the corresponding $D_V$ of RA2, RA3 and RA4 modules also increases to 64, 128 and 256 respectively. In each RA block, $D_K$ is equal to twice $D_V$, and $D_Q$ is equal to $D_K$. As mentioned earlier, the TC module contains 2 RDC blocks, and the corresponding number of convolution kernels $f$ is determined by the size of the dataset $N_d$. Similar to the RA block, the corresponding feature dimension increases by $f$ after each TC module due to residual connections. Fig.6 shows the input and output size of each module when $f$ is 128.

During the network training phase, the learning rate was adjusted to 0.001, number of epochs was set to 50, the number of batches was set to 50, and the Adam optimizer and cross entropy loss function were used. The proportions of the training set, validation set and test set were 70%, 10% and 20% respectively.

# III. MATERIALS AND METHODS

## A. Datasets and Pre-processing

To compare the performance of the proposed method with other works, the Wave_Clus [40] spike bank was used. The database in [40] comprises various average spike waveforms obtained from the neocortex and basal ganglia of humans. To replicate

Table I Comparison of classification performance of the proposed FS-SS and other methods including Wave_Clus, PCA-K , DSD, TDC, MS4 and SC under different datasets and noise levels. Precision, recall and the accuracy are calculated based on Eq. (13), (14) and (15) respectively.

| Dataset group 1 | Noise levels | Proposed | | | Others (Accuracy) | | |
|---|---|---|---|---|---|---|---|
| | | Precision | Recall | Accuracy | Wave_Clus[a] | PCA-K[b] | DSD[c] |
| Easy1 | 0.05 | 99.18 | 99.16 | 99.16 | 98.82 | **99.40** | 99.13 |
| Easy1 | 0.1 | 99.60 | 99.60 | 99.60 | 98.81 | **99.65** | 99.31 |
| Easy1 | 0.15 | 99.64 | 99.64 | **99.64** | 98.72 | 99.45 | 99.13 |
| Easy1 | 0.2 | 99.52 | 99.53 | **99.53** | 98.55 | 99.51 | 98.93 |
| Easy2 | 0.05 | 99.60 | 99.60 | **99.60** | 96.88 | 95.21 | 97.31 |
| Easy2 | 0.1 | 99.89 | 99.89 | **99.89** | 91.62 | 95.18 | 92.40 |
| Easy2 | 0.15 | 99.53 | 99.52 | **99.53** | 91.30 | 96.77 | 92.23 |
| Easy2 | 0.2 | 99.51 | 99.52 | **99.51** | 84.52 | 93.34 | 84.62 |
| Difficult1 | 0.05 | 98.84 | 98.84 | **98.84** | 89.10 | **98.84** | 97.22 |
| Difficult1 | 0.1 | 98.19 | 98.11 | 98.13 | 93.44 | **98.93** | 95.24 |
| Difficult1 | 0.15 | 98.70 | 98.65 | **98.66** | 66.95 | 97.32 | 57.68 |
| Difficult1 | 0.2 | 98.27 | 98.26 | **98.26** | 75.19 | 92.95 | 67.73 |
| Difficult2 | 0.05 | 99.29 | 99.32 | **99.30** | 93.52 | 87.28 | 96.57 |
| Difficult2 | 0.1 | 99.65 | 99.64 | **99.65** | 94.05 | 83.90 | 98.66 |
| Difficult2 | 0.15 | 99.54 | 99.53 | **99.54** | 83.16 | 72.63 | 91.87 |
| Difficult2 | 0.2 | 99.68 | 99.69 | **99.69** | 56.17 | 32.32 | 78.82 |
| **Mean** | | 99.29 | 99.28 | **99.28** | 88.18 | 90.17 | 91.53 |
| **Dataset group 2** | **Noise levels** | **Precision** | **Recall** | **Accuracy** | **TDC[d]** | **MS4[e]** | **SC[f]** |
| Tetrode-C4 | 5μV | 97.70 | 97.66 | **97.67** | 94.84 | 91.02 | 51.47 |
| Neuronexus-C32 | 5μV | 99.20 | 99.21 | **99.20** | 83.31 | 72.80 | 77.58 |
| Neuropixels-C64 | 5μV | 98.37 | 98.37 | **98.37** | 81.25 | 67.12 | 55.75 |

**a)** DWT(four-level Haar wavelet) and Kolmogorov–Smirnov used in Wave_Clus.
**b)** PCA-K uses three principle components in conjunction with K-means algorithm to classify the spikes.
**c)** Deep spike detection (DSD) uses two CNNs to identify the active spike channels and to extract authentic spike data.
**d)** TDC(Tridesclous) employs automatic clustering based on density-peak detection (using PCA-reduced features) combined with template matching.
**e)** MS4(MountainSort4) utilizes a three-stage pipeline: noise-robust spike detection via amplitude thresholding, drift-insensitive feature extraction (using geometric embedding), and scalable clustering with isosplit.
**f)** The core idea of SimpleClust (SC) is to perform basic clustering operations directly on the extracted spike waveforms.

the background noise activity, attenuated spike waveforms selected at random from the data library were incorporated into the generated datasets. There are four datasets in the collected database, each has three spike mean waveforms and provides corresponding spike times and their labels. Besides, the four datasets are categorized according to the different degrees of difficulty (e.g., similarity of spike shape) and the noise levels. The datasets are labeled as C_Easy1_noise, C_Easy2_noise, C_Difficult1_noise, and C_Difficult2_noise, with noise levels represented by standard deviations ($\sigma_N$) of 0.05, 0.1, 0.15, and 0.2. The terms "Easy" and "Difficult" refer to the similarity index between spike shapes in each dataset. Easy1 has also additional noise levels of 0.25, 0.3, 0.35 and 0.4 for further spike sorting performance analysis. Based on the time information for each spike, a sampling window consists of 66 samples was generated to extract the spike waveforms to examine the classification performance of the proposed FS-SS framework.

In addition to the Wave_Clus datasets (Datasets Group 1), a new set of synthetic datasets, referred to as Datasets Group 2, is generated with the SpikeInterface [41] to further evaluate the sike sorting algorithms. Datasets Group 2 comprises Tetrode_C4 utilizing a tetrode probe with 4 channels; Neuronexus_C32 simulating a Neuronexus probe with 32 channels and Neuropixels_C64 representing a Neuropixels probe with 64 channels. The number of spike waveforms is set to six, which means that the Group 2 dataset is a six-class classification problem. The sampling rate is set to 30 kHz, the firing rate for all units is 15 Hz and a refractory period of 2 ms was applied during the data generation. The noise level across all channels and datasets was standardized at 5 μV. The total recording duration for Tetrode_C4 was 600 seconds, while Neuronexus_C32 and Neuropixels_C64 each had a generation time of 100 seconds. These diverse datasets allow for a comprehensive assessment of the FS-SS framework's adaptability and performance across varying channel counts and probe architectures.

Before the spike waveforms are fed to the network for classification, a simple min-max normalization is performed on the spike data.

### B. Evaluation Methods

Following the works in [1], [41], standard classification metrics including $precision$, $recall$, and $accuracy$ were used to evaluate the evaluate the sorting performance. In the metrics, true positive spikes (TPs) represents correctly identified spikes from the target neuron; false positive spikes (FPs) indicates non-target spikes erroneously assigned to the target neuron; false negative spikes (FNs) refers to the actual target spikes that were misclassified as belonging to other neurons and true negative spikes (TNs) are the non-target spikes that were correctly excluded from the target neuron's classification. $Precision$ is determined by dividing the number of true positive spikes (TPs) by the total number of all predicted positive spikes, including true positive spikes (TPs) and false positive spikes (FPs):

$$Precision = \frac{TPs}{TPs + FPs} \tag{13}$$

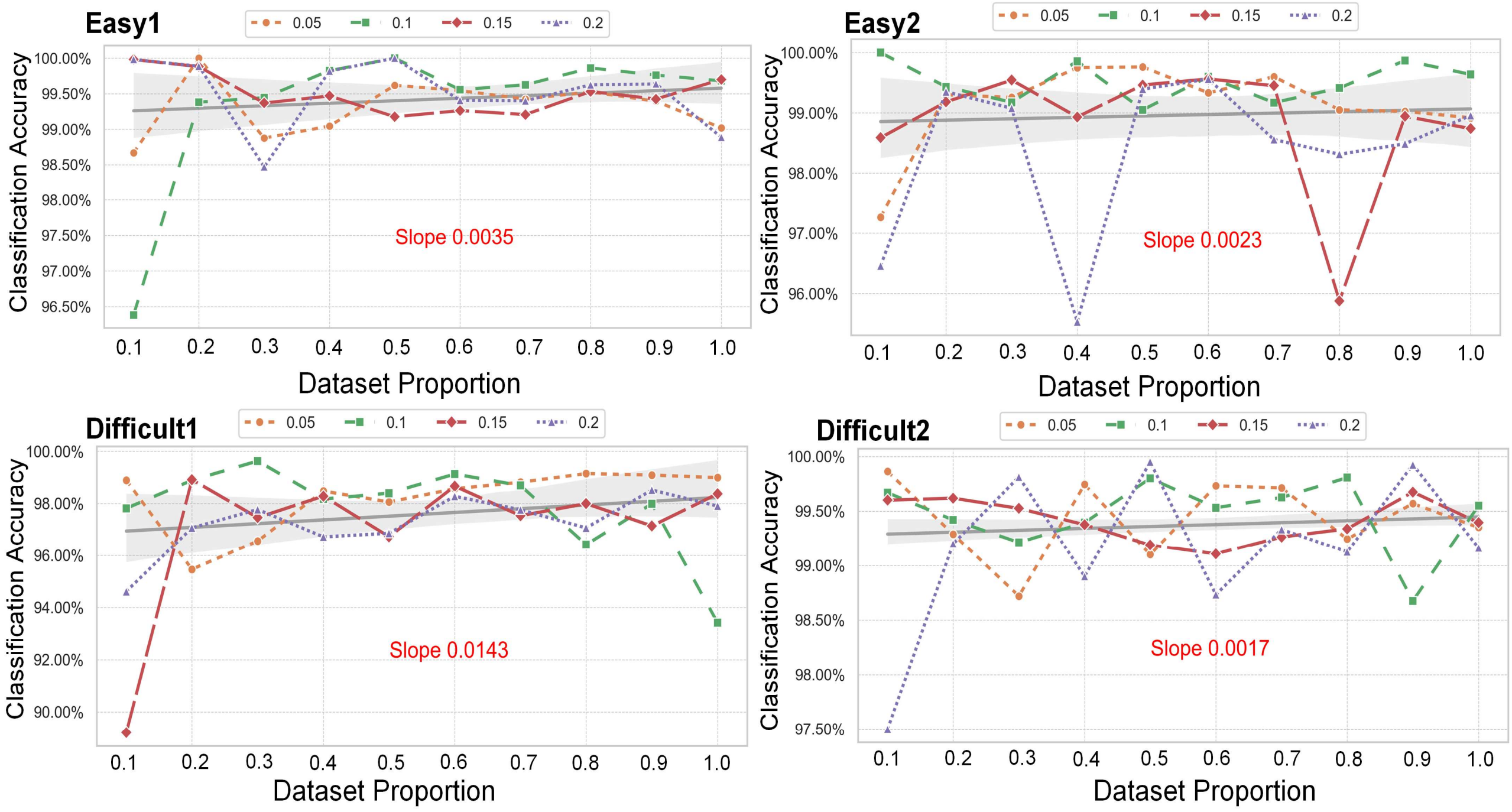

Fig.7. Classification accuracy of Easy1, Easy2, Difficult1, and Difficult2 datasets from Dataset Group 1 at different standard deviations ($\sigma_N$) of 0.01, 0.05, 0.15, and 0.2 versus dataset proportions ranging from 10% to 100%. The shaded area represents the mean accuracy that corresponds to more than >95% confidence level in the datasets. Slope in each figure is calculated from curve fitting (gray straight line), which indicates the impact of increasing the dataset proportion on the classification accuracy. Overall, FS-SS represents an outstanding robustness to the dataset proportion in the adaptation process.

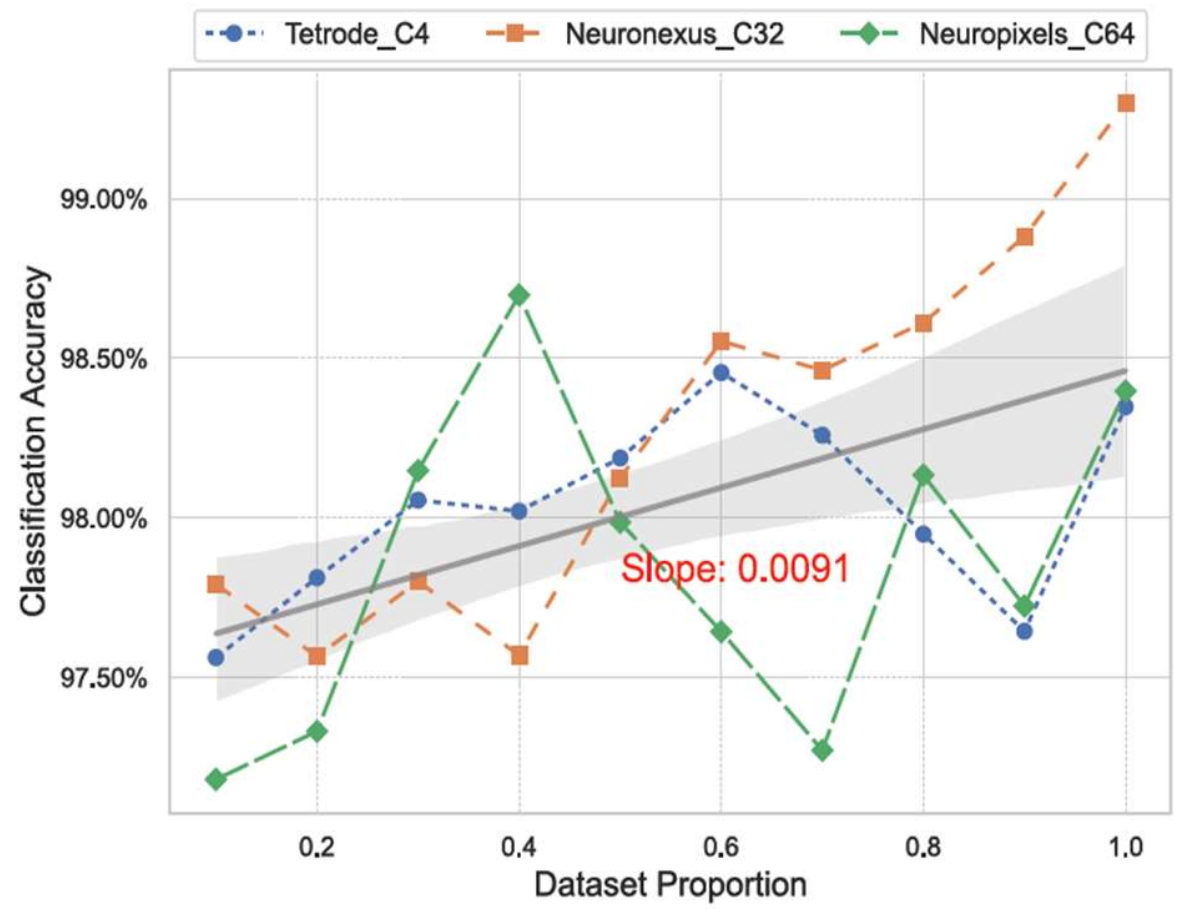

Fig.8 Classification Accuracy with Different Dataset Proportions for Datasets Group 2. The figure displays the classification accuracy of the FS-SS model as the proportion of the training dataset varies from 0.1 to 1.0 for Tetrode_C4, Neuronexus_C32, and Neuropixels_C64 datasets. The grey line represents the overall trend.

*Recall* is determined by dividing the number of true positive spikes (TPs) by the number of all actual positive spikes, including true positive spikes (TPs) and false negative spikes (FNs) expressed as:

$$Recall = \frac{TPs}{TPs + FNs} \tag{14}$$

*Accuracy* indicates the proportion of correctly classified spikes, expressed as the sum of true positive (TPs) and true negative (TNs) divided by the number of all spikes, including true positive spikes (TPs), true negative spikes (TNs), false positive spikes (FPs) and false negative spikes (FNs), expressed as:

$$Accuracy = \frac{TPs + TNs}{TPs + FPs + FNs + TNs} \tag{15}$$

Table II Dropout rate, number of convolution kernels and model size for different data proportions.

| Proportion | Dropout rate | Kernels numbers | Model Size |
|---|---|---|---|
| 10% (0.1) | 0.5 | 8 | 967,502 |
| 20% (0.2) | 0.45 | 16 | 1,238,070 |
| 30% (0.3) | 0.41 | 16 | 1,238,070 |
| 40% (0.4) | 0.36 | 32 | 1,804,550 |
| 50% (0.5) | 0.32 | 32 | 1,804,550 |
| 60% (0.6) | 0.27 | 32 | 1,804,550 |
| 70% (0.7) | 0.23 | 32 | 1,804,550 |
| 80% (0.8) | 0.18 | 64 | 3,038,886 |
| 90% (0.9) | 0.14 | 64 | 3,038,886 |
| 100% (1) | 0.1 | 64 | 3,038,886 |

## IV. Experiments and Results

In this section, the proposed FS-SS framework is applied to the spike classification task. First, the performance of the FS-SS framework is compared with existing state-of-the-art algorithms. Then, to explore the adaptability of the model, explained dataset in Section III are divided into different portions to explore the performance of the FS-SS under different dataset sizes. Furthermore, attention visualization in various scenarios are presented in order to understand the feature extraction capability and interpretability of the proposed FS-SS network. Finally, ablation experiments were performed to explore the impact of each module on the overall performance.

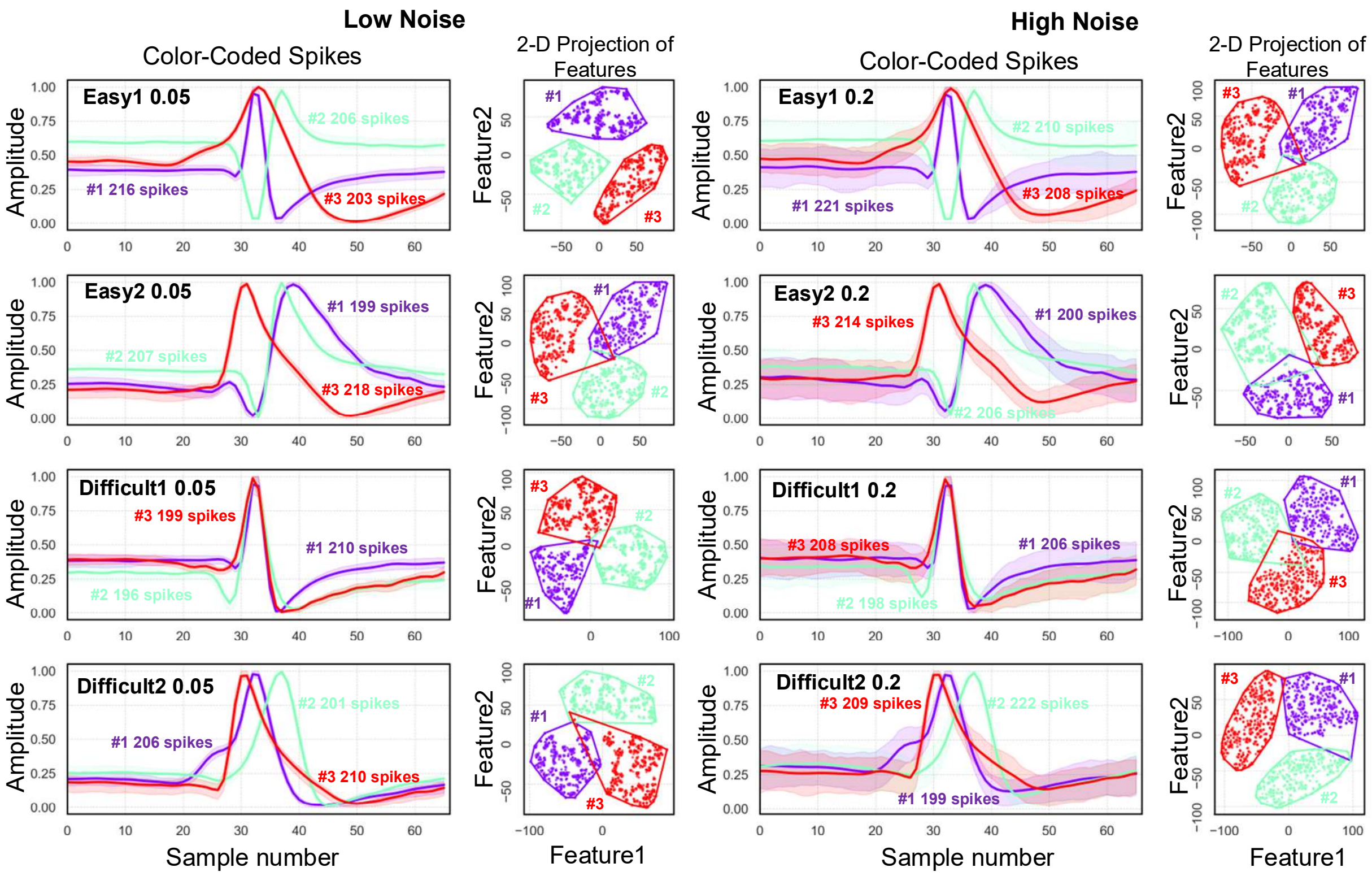

Fig.9. Alignment and feature visualization of spike waveforms of different datasets for $\sigma_N$ =0.05 (left column) and $\sigma_N$=0.2 (right column). Under each column, color-coded spike waveforms are superimposed based on the ground-truth. The shaded part represents the 60% data distribution range. Annotated numbers next to the spikes show the number of predicted and aligned waveforms in each class. The scatter plot under left or right columns shows the visualization result of the 128-dimensional features after the embedding module using t-SNE dimensionality reduction. Each color represents a different class, and the envelope shows the boundary of the corresponding cluster.

### *A. Evaluation of Classification Performance*

The sorting performance of the proposed FS-SS framework and other state-of-the-arts including Wave_Clus [40], combination of PCA and K-means (PCA-K) [42], deep spike detection (DSD) [3], Tridesclous (TDC) [43], Mountainsort4 (MS4) [44], Simpleclust(SC) [45] are shown in Table I. Algorithms in [3, 40, 42-45] are briefly explained in the footnote of Table I. All three explained performance metrics in Section III. B, $Precision$, $Recall$ and $Accuracy$ are evaluated and reported in Table I for the considered spike sorting methods. To ensure the fairness of the comparisons, the mean sorting accuracy of seven epochs is calculated and two epochs with the highest and lowest performance values are removed (i.e. removing deviated values). Therefore, the performance metrics listed in Table I are the average of 5 epochs.

The comparison of FS-SS, Wave_Clus [40], PCA-K [42], and deep DSD [3] is performed across all spike shapes in "Easy" and "Difficult" datasets and the noise levels (0.05, 0.1, 0.15 and 0.2). It is observed from Table I that in Easy2_0.05, the proposed FS-SS reaches $Precision = 99.18\%$ , a $Recall$ of 99.16% and 99.16% $Accuracy$. This is lower than the PCA_K method, but higher than the Wave_Clus and DSD methods. When the noise standard deviation ($\sigma_N$) is 0.1, the accuracy is still slightly lower than the PCA-K method. But when the noise level increases to 0.15 and 0.2, the accuracy of the proposed FS-SS is higher reaching to 99.64% and 99.53% respectively. In the other datasets, the FS-SS achieves higher sorting accuracy except the PCA_K method on the Difficult1 dataset ($\sigma_N$=0.1).

For Dataset Group 2 datasets with multiple channels, the FS-SS achieves significant advantages especially in the case of multiple channels (Neuronexus_C32 and Neuropixels_C64). Compared with TDC and MS4 algorithms, for the Tetrode_C4 dataset with four channels, the lead of FS-SS is 3% to 8%. Since SC simply clusters features of the spike waveforms, its accuracy is greatly reduced to only 51.47%. On the 32-channel Neuronexus_C32 and 64-channel Neuropixels_C64 datasets, FS-SS's lead continues to expand. However, it should be noted that the algorithms used for comparison are integrated into Spikeinterface, involving multiple processes, each of which has corresponding parameter settings. The default parameters were used in the experiment. In short, it is observed that FS-SS is in an absolute leading position compared to other algorithms.

This demonstrates the FS-SS network's ability to accurately identify the most informative features for sorting step, and its extreme robustness to spike similarity and noise level.

### *B. Learning with Different Dataset Proportions*

As explained in Section II. B , linear interpolation was used to obtain optimal number of convolution kernels in feature extraction and to limit the size of the entire model to avoid unnecessary computations. Therefore, this section further evaluates the performance of scale-adaptive FS-SS model by using of 10 different data proportions ranging from 10% (or 0.1) to 100% (or 1). Data proportions refer to the number of the data samples used to simulate and verify the adaptation capability and classification performance of the FS-SS framework. Table II lists the total number of model parameters in different dataset proportions, as well as the corresponding hyperparameter information such as the number of convolution kernels and dropout rate. Fig.7 shows the classification performance of the FS-SS model for different training dataset proportions and different noise levels (taking the last 12 epochs and removing the maximum and minimum values).

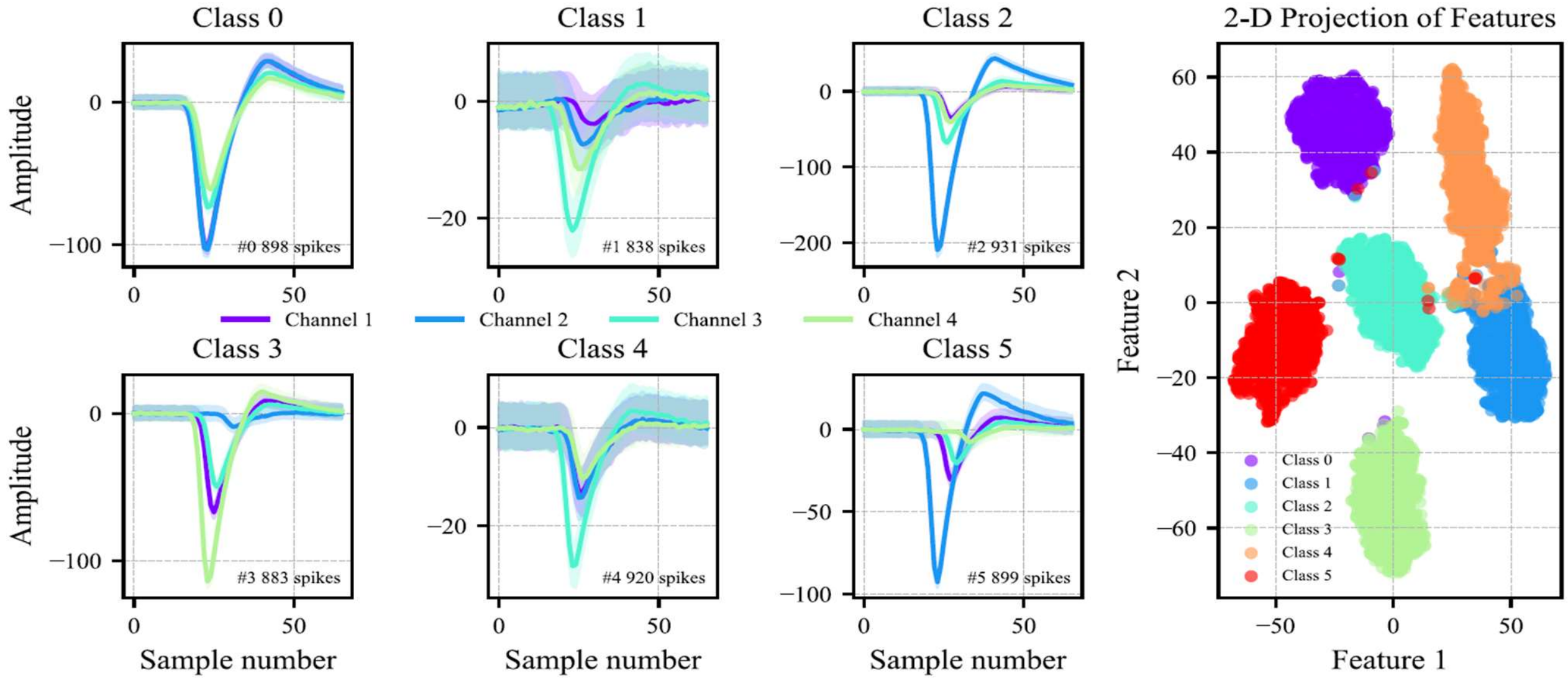

Fig.10 Spike alignment and 2-D t-SNE projection of features for Tetrode_C4 from Datasets Group 2. Class 0-5shows the aligned spike waveforms across four different channels (Channel 1-4) for six distinct classes from Tetrode_C4. The right panel illustrates the 2-D t-SNE projection of the extracted features, where each color represents a different spike class, demonstrating clear clustering and separability.

From the overall trend in Fig.7, the FS-SS model shows a high classification accuracy in most cases, with slight fluctuations on small dataset proportions. As the dataset proportion continues to increase, the accuracy of the model tends to stabilize. Especially in the case of low noise ($\sigma_N$=0.05 and $\sigma_N$=0.1), the accuracy remains high across different data proportions showing extremely high robustness. In the case of high noise ($\sigma_N$=0.15 and $\sigma_N$=0.2), although the accuracy fluctuates to a certain extent, the overall mean accuracy remains high (i.e. minor fluctuations). This shows that the FS-SS has extremely fast adaptability to the recording channels changes over time even when 10% of data is used for training purposes. Assuming that there are originally 3514 spike waveforms for training and testing in Easy1_0.05. 10% of the spikes for training and testing results in 98.7% accuracy. The performance of the FS-SS network evolves quickly by employing meta-learning mechanism and adaptive hyperparameters even under small training data conditions.

A linear fit was performed to further understand the mean accuracy versus dataset proportions. The slope of the fit was used to quantify the impact of the dataset proportion on the model performance. The gray shaded areas shown in Fig.7 represents an incremental trend and the calculated slopes are all positive which indicate the increase in the proportion of the datasets resulted in accuracy improvement in. For example, in Difficult1, the overall mean accuracy is dropped by 1.43% from 0.0143, when dataset proportion is reduced by 90%. This results in the total number of FS-SS parameters reduction from 3M to 0.9M as shown in Table II (e.g. when the number of convolution kernels is reduced from 64 to 8). In Easy1, Easy2 and Difficult2 datasets, reducing the dataset proportion by 90% only reduces the accuracy by 0.35%, 0.23%, and 0.17%.

The classification performance of Dataset Group 2 is also studied when different data ratios used to verify the adaptability of the FS-SS in multi-channel cases. Fig.8 illustrates that for all three datasets, the classification accuracy generally improves as the data proportion for training of the FS-SS is increased. The consistent upward trend is quantified by an overall slope of 0.0091 which demonstrates the model's ability to leverage more data for enhanced performance. Specifically, the Neuronexus_C32 dataset shows a particularly smooth and consistent increase in accuracy, reaching its peak at the full dataset proportion. While Tetrode_C4 and Neuropixels_C64 exhibit some fluctuations, their overall performance also benefits from larger training subsets. Even with smaller dataset proportions (e.g., 0.1 or 0.2), the FS-SS framework maintains over 97% accuracy sorting accuracy, highlighting its efficiency in learning from limited samples and its adaptability to different recording environments.

The above analysis shows that the FS-SS model can successfully reduce the total number of parameters of the entire network by reducing the number of convolution kernels when facing a small dataset, without decisively affecting the overall classification performance. This is associated to two reasons: first, dynamically adjusting the number of convolution kernels and the size of the dropout rate can effectively alleviate overfitting and improve the performance of the model on small datasets. Second, the FS-SS model utilizes meta-learning which is designed for adaptations using small datasets. By training on multiple related tasks, the FS-SS network can quickly learn key features when faced with new tasks. This proves the superior performance of the FS-SS network on spike classification with real-time adaptability.

This capability is directly translatable to practical BMI scenarios, where obtaining large amounts of labeled data for new neurons or subjects is often challenging and time-consuming. The FS-SS method significantly reduces the calibration latency, accelerating the deployment and adaptation of BMI systems.

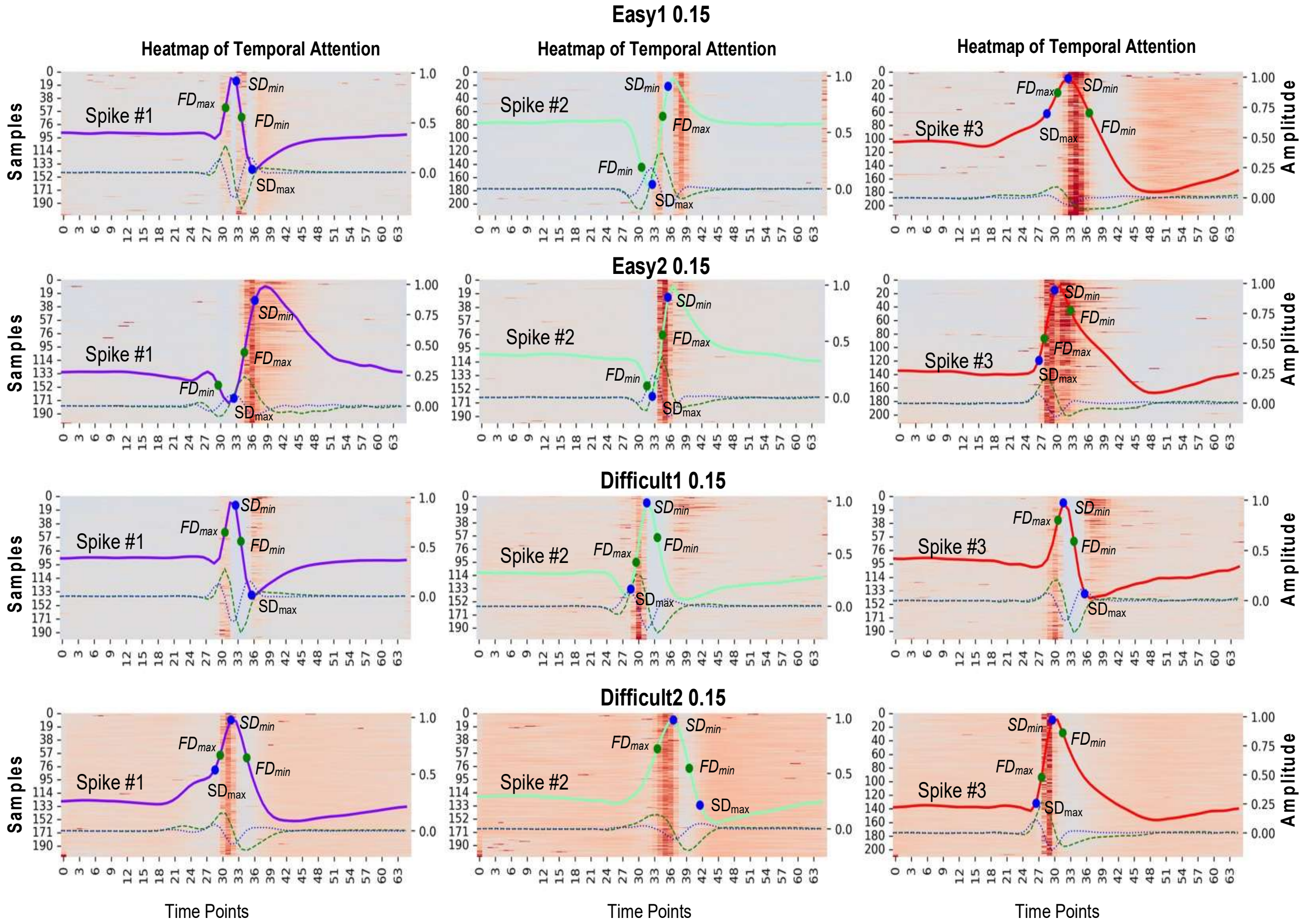

Fig.11 Attention heatmaps of three sike waveforms in different datasets in Easy1_0.15 (first row), Easy2_0.15 (second row), Difficult1_0.15 (third row) and Difficult1_0.15 (fourth row). The horizontal axis of each sub-graph represents the time point, and the vertical axis represents the sample number. The darker color reflects the degree of attention in the RA module to different time points. Three waveforms are also superimposed on the heatmaps: a) the mean of the spike waveform according to the predicted label, b) the green dotted line that represents its corresponding first-order derivative (FD), and the blue dotted line represents the second-order derivative (SD). The key features are marked including $FD_{max}$ (maximum value of the first derivative), $FD_{min}$ (minimum value of the first derivative), $SD_{max}$ (maximum value of the second derivative), and $SD_{min}$ (minimum value of the second derivative).

### C. Spike Alignment and Feature Visualization

In order to further understand the feature extraction ability of the FS-SS, Fig.9 shows two-dimensional (2D) scatter plots of the embedding module after t-SNE (t-Distributed Stochastic Neighbor Embedding) dimension reduction [46]. The left and right columns in Fig.9 represent low ($\sigma_N$=0.05) and high ($\sigma_N$=0.2) noise datasets from Dataset Group 1. Each row in Fig.9 also represents the type of dataset. The test spike waveforms are color-coded according to the ground-truth to better distinguish between the neurons. Aligned color-coded mean spike waveforms along with their shaded fluctuation range are shown for each dataset.

It is observed from Fig.9 that, whether for the Easy dataset with lower similarity or the Difficult dataset with higher similarity, the mean waveforms are clear and identifiable which shows the proposed FS-SS network strong ability to extract time domain features of different spikes. Noisy conditions introduce a wider shaded area, however the superimposed classified waveforms still show an acceptable degree of separation. This shows that the noise has very little impact on the aligned spikes after classification, and the proposed FS-SS grasps the key features and maintains stable performance in the face of high noise.

The 2D scatter plot on the right shows the t-SNE dimensionality reduction result of the embedding module after label prediction in the test set. It is observed that the feature points after dimensionality reduction show robust classification, and the boundaries between clusters corresponding to different classes are clear. Even in the case of high noise, there will be no high overlaps between different 2D cluster projection borders which shows that the FS-SS model can effectively encode features and so to distinguish features of different categories.

Similar to the analysis performed for dataset group 1, spike alignment and 2D projection of features using t-SNE for the Tetrode_C4 dataset is presented in Fig.10. The aligned spike waveforms for each class (classes 0-5) from different channels (channels 1-4) demonstrate the different morphologies captured by the framework. The corresponding t-SNE feature visualization demonstrates the effectiveness of the FS-SS framework in extracting discriminative features, resulting in clear clusters for each spike class in the 2D embedding space. This further demonstrates the robustness and accuracy of the proposed method on a variety of challenging datasets.

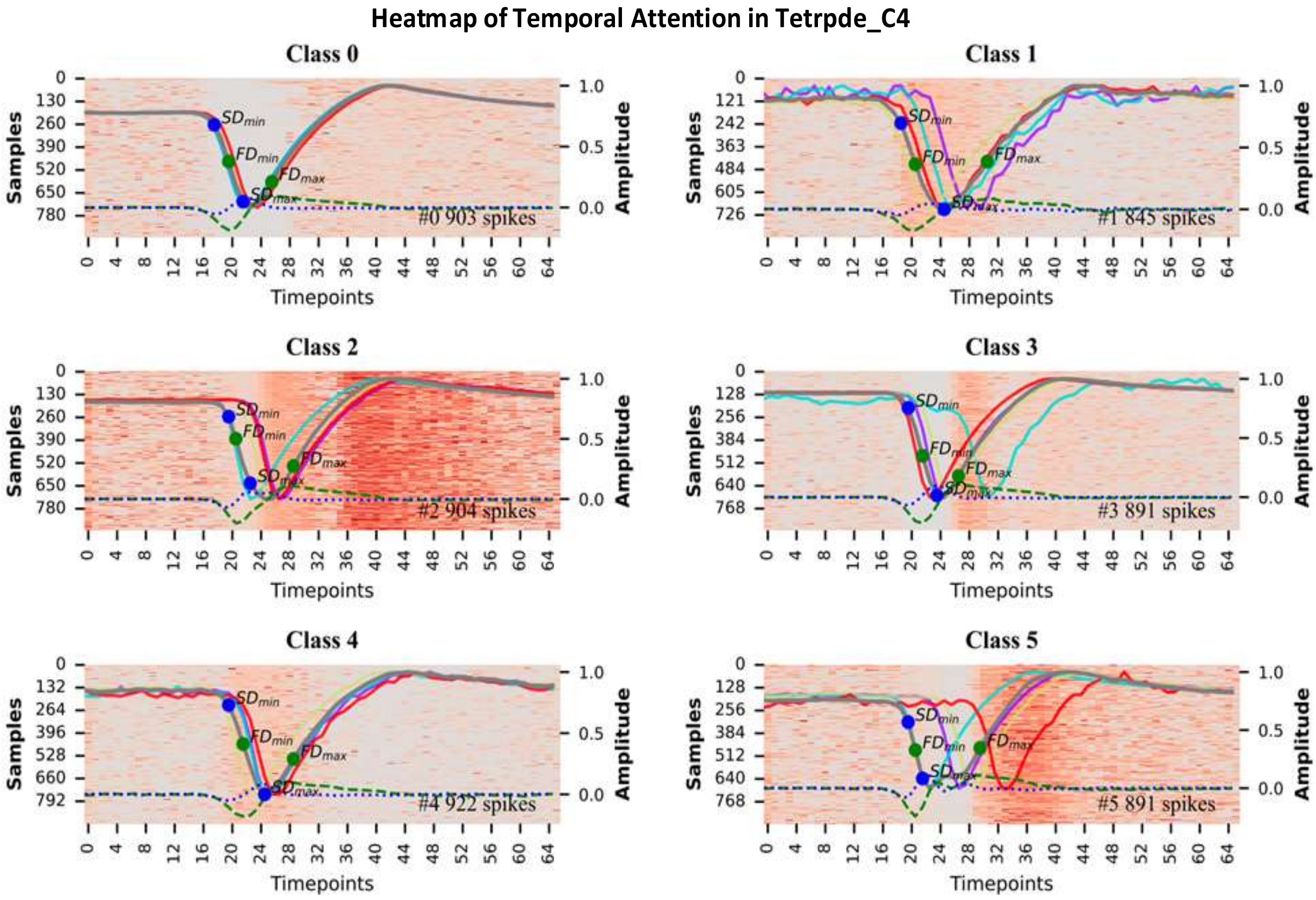

Fig.12 Heatmap of temporal attention for Tetrode_C4. Each sub-figure presents the temporal attention heatmap for a specific spike class (Class 0 to Class 5) from Tetrode_C4. The attention intensity, indicated by color, highlights the locations of important features (e.g., $SD_{min}$, $FD_{min}$, $FD_{max}$, $SD_{max}$) on a spike waveform that the FS-SS model focuses on for classification.

Overall, evaluations against various noise levels highlights the strong robustness of FS-SS. This inherent resilience to background noise and similar spike waveforms ensures high classification accuracy in real-world neural signals, which are frequently noisy and variable, leading to more reliable and stable BMI control.

## *D. Visualization of Attention Results*

This section illustrates the time domain feature heatmap of different spike waveforms in the embedding module since this has a significant impact on the overall classification performance. Visualization of the attention distributions in different test datasets at a noise standard deviation of ($\sigma_N$=0.15) are shown in Fig.11, for example the first row in Fig.11 corresponds to the Easy1_0.15 dataset.

Fig.11 also superimposes the spike waveform, its first-order derivative (FD) by green dashed line and its second-order derivative (SD) using blue dotted line. $FD_{max}$, $FD_{min}$, $SD_{max}$, $SD_{min}$ representing maximum and minimum of FD and SD are marked in each figure for better understanding of attention heatmaps. It is observed from the top row in Fig. 9 that for the Easy1_0.15, the attention heatmap mainly focuses on the $FD_{max}$ and $FD_{min}$ points of spike #1. There is also a part of attention distributed after $SD_{max}$. For spike #2, part of the attention is distributed in the rising phase near $FD_{max}$ which represents the phase with the fastest voltage rise (i.e. maximum slope). Another part of the attention is distributed in the falling phase after $SD_{min}$, suggesting that the model considers this phase as the key morphological feature in spike #2 in Easy1_0.15. For spike #3 in Easy1_0.15, the attention distribution is slightly different, with one feature concentrated between $SD_{min}$ and $FD_{min}$ and the other part distributed after $FD_{min}$. This indicates that the proposed model believes the slow transition after $FD_{min}$ is the key feature to distinguish spike #3 which is consistent with observations. These results strongly indicate that the proposed FS-SS network is not completely a "black box", on the contrary, it has a fairly strong electrophysiological interpretability.

In Easy2_0.15 dataset, the distribution of attention is mainly concentrated on the rising phase of the spike aligned with $FD_{max}$ and after the $SD_{min}$. In the Difficult1_0.15 dataset with higher waveform similarity, there is no higher attention distribution near $FD_{min}$ (i.e. the region with the fastest voltage drop) which indicates that the FS-SS believes this segment cannot effectively distinguish the three spike waveforms. More specifically, part of the attention is distributed around $FD_{max}$, where there is a fast voltage rise. Another part of the attention is distributed around $SD_{max}$ and after that. This shows that the model observes more subtle differences in morphological features and places more attention on these two areas. The attention distribution corresponding to the Difficult2_0.15 dataset is almost similar to Difficult1_0.15, with more attention near $FD_{max}$. There is not so much attention distribution in the falling stage around $FD_{min}$.

To provide deeper insights into how the FS-SS model processes different spike classes, the temporal attention distribution for Tetrode_C4 from Datasets Group 2 is shown in Fig.12. Each subplot in Fig.12 displays the heatmap of temporal attention for a specific spike class, superimposed on a representative set of spike waveforms. The FS-SS consistently focuses its attention on the crucial features of the spike waveforms, such as the initial depolarization, peak amplitude ($SD_{min}$ and $FD_{min}$), and repolarization phases ($SD_{max}$ and $FD_{max}$). The attention distribution of some classes is not concentrated near the extreme points as the network believes the extrema features cannot distinguish this class from other classes, such as Class 0. In short, the proposed model provides a fairly strong electrophysiological interpretation through attention in distinguishing the most informative features in different spikes, and can dynamically adjust attention according to the characteristics of different morphologies. The dynamic attention distribution adjustment is not only used to assist in observing direct or indirect features, but also demonstrates the superior performance of the FS-SS network in spike sorting.

### *E. Module Ablation and Comparison*

To systematically evaluate the contribution of each key component within the proposed FS-SS framework, module ablation is studied. This involved freezing the parameters of specific module to assess its individual impact on the overall performance of the FS-SS. These experiments were performed on the Difficult2 dataset from Datasets Group 1 and all datasets from Datasets Group 2 including Tetrode_C4, Neuronexus_C32, and Neuropixels_C64. The results of the ablation studies are summarized in Fig.13.

The results in Fig.13 show that after freezing the embedding module, the classification accuracy of the model dropped significantly which highlights the significance of this module in capturing and representing the characteristics of the original data. It is worth noting that on the Neuronexus_C32 dataset, the average drop in accuracy was as high as 20.5%, indicating that the embedding layer is essential for processing the complex or high-dimensional features of this particular dataset. On the Difficult1 dataset, the mean accuracy dropped by about 3% under different noise levels when embedding module is removed.

The ablation experiment results of the RA (residual attention) module show that its impact on the model accuracy is not very significant. On the Difficult2 dataset, the P value is relatively small (0.001 to 0.01), and on the Neuropixels_C64 dataset, the P value is slightly increased (between 0.01 and 0.05). On the Tetrode_C4 and Neuronexus_C32 datasets, removing the RA module did not even produce statistical significance ($P>0.05$), which indicates that the direct contribution of the RA module to the model performance in these specific scenarios has not reached a statistically significant level. This is because the embedding module already captured the discriminative features of the spike waveforms, making the role of the RA module redundant. In addition, the characteristics of the dataset may also limit the potential advantages of the attention mechanism. The RA module may still play a subtle role that is difficult to quantify in improving the robustness or generalization ability of the FS-SS model.

For the TC module, the accuracy of the model also diminishes following the freezing of this module; however, the effect is less severe compared to that of the embedding module. The accuracy of the Neuropixels_C64 dataset drops by 3.4%, which may imply that the TC module is highly relevant for processing more sophisticated or complex spike patterns in this dataset. The Neuronexus_C32 dataset experiences the smallest decline in accuracy, approximately 1.2%. This could indicate that the temporal features identified by the TC module have a limited effect on the ultimate classification outcomes, or that the Embedding module has successfully captured the majority of the essential information. The implementation of the TC module can still yield enhancements in performance, particularly in datasets that require the capture of more complex temporal dynamics, and its significance should not be overlooked. In summary, the embedding module has been proven to be the most critical component of the model and has a decisive impact on the overall classification performance. The TC module has improved the model's temporal feature learning ability to varying degrees. Although the RA module did not have a statistically significant impact on all data sets in this ablation experiment, its potential auxiliary role and contribution to the model's generalization ability are still worth further exploration.

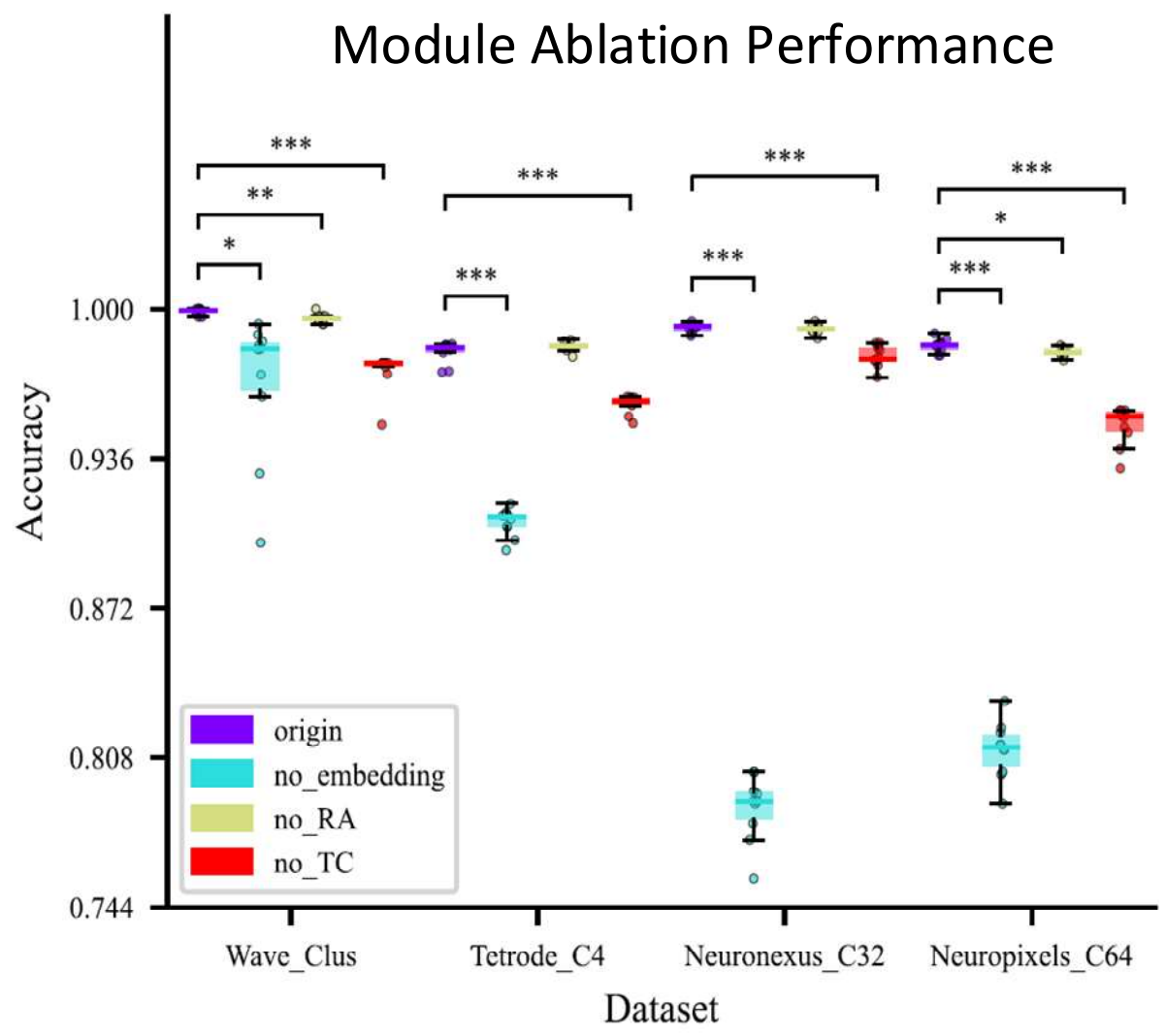

Fig.13 Module ablation performance. This figure shows the accuracy of the FS-SS model under different module ablation conditions on different datasets. "Origin" represents the FS-SS. "No_embedding" represents the FS-SS without the embedding module, "no_RA" indicates FS-SS when RA module is removed, and "no_TC" shows the exclusion of the TC module. The datasets comprise Wave_Clus (Difficult2), Tetrode_C4, Neuronexus_C32, and Neuropixels_C64. *, ** and *** represent significant p-values less than 0.05, 0.01 and 0.001 respectively.

## V. Conclusion

This paper proposed a FS-SS framework based on meta-learning and attention mechanism which is suitable for implantable brain processing (referred as spike sorting), where collecting training data is an extremely challenging task. The proposed framework consists of crucial building blocks including an embedding module, residual attention (RA) module and residual dilated convolution (RDC) module. Combination of these key modules ensure quick generalization of learning with only two-shot from each class by building upon the previous experiences in the learning phase, and also extraction of the most informative features in the classification stage. The results show that the proposed framework achieves leading performance on the public dataset. Specifically, when using all samples, FS-SS achieved an average precision of 99.29%, an average recall of 99.28%, and an average classification accuracy of 99.28%. Even on sub-datasets with high noise or high similarity, the proposed framework shows robust outcomes. In addition, by designing hyperparameters that can be dynamically adjusted according to the number of training data, the parameter size of the model is adaptive. When the number of samples is reduced by 90%, the model size becomes 31.8% of the original, while the average accuracy only decreases by 0.55%. This makes it highly suitable for implantable on-chip spike sorting algorithms with extremely high performance and power requirements. Finally, the t-SNE results of the embedding layer and the visualization of the time-domain attention results prove that the framework has a certain physiological explanation rationality and provides more trust in terms of sorting robustness.

In summary, the innovation of the FS-SS framework extends beyond achieving high classification accuracy. It provides a spike classification solution that is inherently adaptive, computationally efficient, robust to real-world noise, and capable of rapid generalization. These are all key properties for the successful and scalable implementation of brain-computer interfaces. The FS-SS's ability to address these practical challenges significantly reinforces its relevance and contribution to the BMI community, paving the way for more robust and user-friendly BMI systems. In future work, the FS-SS will be further improved and actual tests will be conducted using in-vivo datasets.